\documentclass[aps,prl,floatfix,twocolumn,superscriptaddress]{revtex4}

\usepackage{graphicx}
\usepackage{amssymb}
\usepackage{amsmath}

\begin{document}

\title{An atomic colour superfluid via three-body loss}

\date{\today}

\author{A. Kantian}
\affiliation{Institute
for Theoretical Physics, University of Innsbruck, A-6020 Innsbruck, Austria\\ and Institute for Quantum Optics and Quantum Information of the
Austrian Academy of Sciences, A-6020 Innsbruck, Austria}
\author{M. Dalmonte}
\affiliation{Institute
for Theoretical Physics, University of Innsbruck, A-6020 Innsbruck, Austria\\ and Institute for Quantum Optics and Quantum Information of the
Austrian Academy of Sciences, A-6020 Innsbruck, Austria}
\affiliation{Dipartimento di Fisica dell'Universitˆ di Bologna and INFN, via Irnerio 46, 40127 Bologna, Italy}
\author{S. Diehl}
\affiliation{Institute
for Theoretical Physics, University of Innsbruck, A-6020 Innsbruck, Austria\\ and Institute for Quantum Optics and Quantum Information of the
Austrian Academy of Sciences, A-6020 Innsbruck, Austria}
\author{W. Hofstetter}
\affiliation{Institute for Theoretical Physics, Johann Wolfgang Goethe-Universit\"at, Frankfurt, Germany}
\author{P. Zoller}
\author{A. J. Daley}
\affiliation{Institute
for Theoretical Physics, University of Innsbruck, A-6020 Innsbruck, Austria\\ and Institute for Quantum Optics and Quantum Information of the
Austrian Academy of Sciences, A-6020 Innsbruck, Austria}

\begin{abstract}
Large three-body losses in a three-component Fermi gas confined in
an optical lattice can prevent the occupation of a lattice site by
three atoms. This effective constraint not only gives rise to a
suppression of actual three-body loss, but stabilises BCS pairing
phases by suppressing the formation of trions. We study the effects
of the constraint using bosonisation and density matrix
renormalisation group techniques (DMRG). We discuss the case of
lithium experiments, and study the dissipative dynamics including
loss using time-dependent DMRG with quantum trajectories methods.

\end{abstract}

 \maketitle Recent developments in the experimental
control of degenerate Fermi gases with cold atoms \cite{fermiexp}
has paved the way for the study of three-component Fermi mixtures
\cite{lithium}. For attractive two-body interactions, these systems
offer a chance to observe competition between an atomic colour
superfluid phase with BCS pairing and a trionic phase (with three
particles on the same site) \cite{hofstetter,dmrg3,leggett,demler}.
A key feature of current experiments, though (e.g., with lithium),
is the large three-body loss rate observed in these mixtures. Here
we discuss how these losses, which are normally undesirable, can
give rise to an effective three-body hard-core constraint
\cite{daleyloss, pfaffianloss} when the gas is loaded into an
optical lattice \cite{opticallattices}. Not only does this suppress
actual loss events, but it also acts to suppress trion formation,
thus stabilising pairing phases such as the atomic colour
superfluid. The blockade mechanism is related to the continuous
quantum Zeno effect, and has been demonstrated for two-body
interactions in an experiment with Feshbach molecules
\cite{moleculeloss1,moleculeloss2}. The constraint has  particularly
striking consequences in 1D, where in the absence of such a
constraint there is no atomic colour superfluid  (ACS) phase, but
rather competition between a charge-density wave (CDW) and a phase
with symmetric (on-site) trions (ST) \cite{dmrg3} (see Fig.~1a). We
show that a constraint prevents ST formation and produces an ACS
phase, which competes with a CDW and off-site trions (OT) (see
Fig.~1b).

\begin{figure}[t]{
\centering
  {\includegraphics[width=8cm]{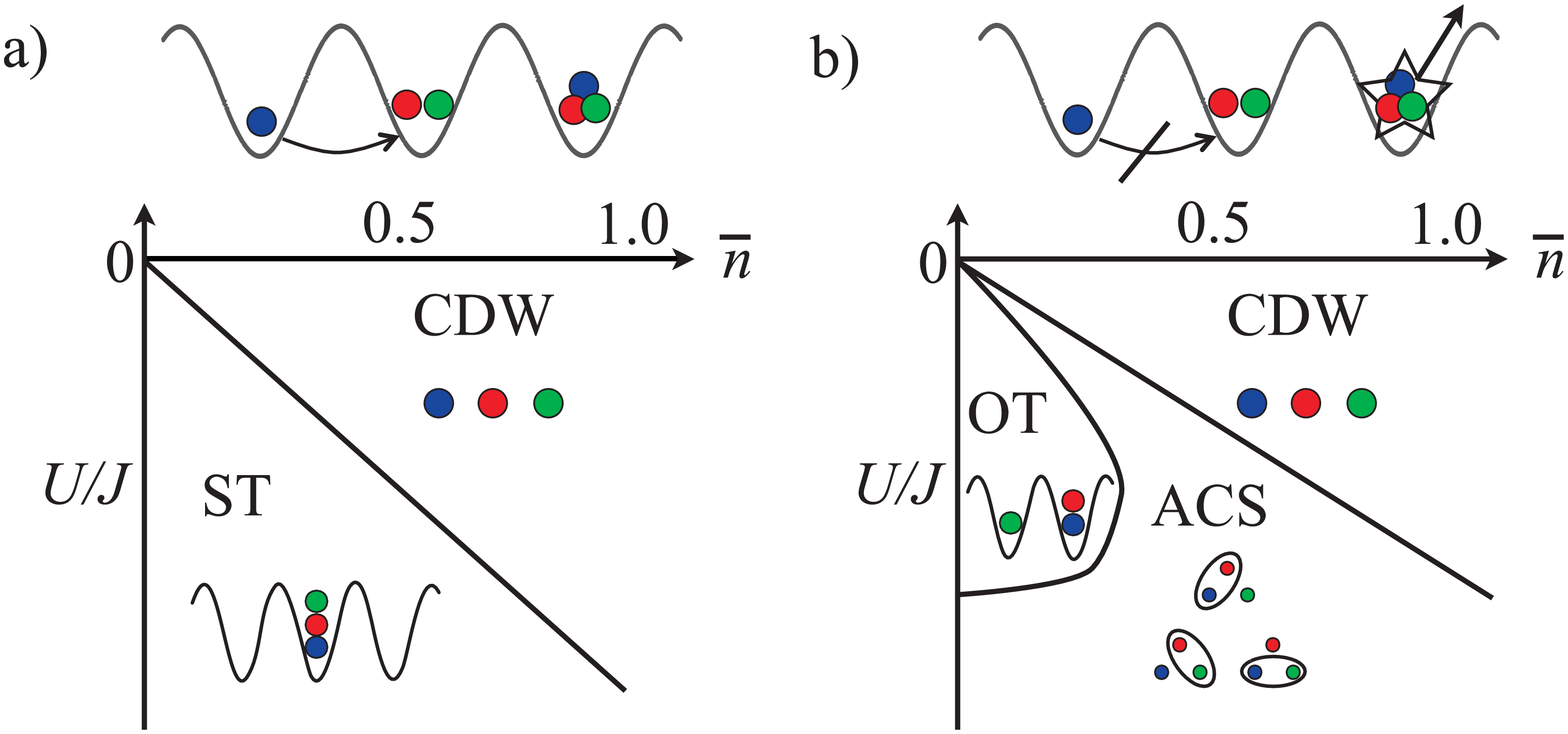}}
\caption{\small{Qualitative phase diagram for attractive
interactions $U<0$ and equal populations ${\bar n} /3$ of each
component in a three-component 1D Fermi gas. These are shown in the
SU(3) symmetric case (where all pairwise interactions between
different components are of equal strength), a) without and b) with
a three-body hard-core constraint arising from three-body loss. The
unconstrained case is characterised by competition between symmetric
(on-site) trions (ST) and a charge-density wave (CDW). The hard-core
constraint suppresses trion formation, stabilising BCS pairing in an
atomic colour superfluid (ACS), which competes with a CDW and
off-site trions (OT).}}
 \label{phased}}
\end{figure}

A three component Fermi gas in the lowest band of an optical lattice
is described by the Hamiltonian ($\hbar=1$)
\begin{equation}
H_U=-\sum_{\langle i,j\rangle, \sigma}J_{\sigma}(c^{\dagger}_{i,\sigma}c_{j,\sigma}+\textit{h.c.})+\sum_{i, \sigma}{U}_{\sigma}m_{i,\sigma}m_{i,\sigma+1},
\end{equation}
where $\langle i,j \rangle$ denotes a sum over neighbouring sites,
$c^{\dagger}_{i,\sigma},c_{i,\sigma}$ are fermionic operators with a
species index $\sigma=1,2,3$,
$m_{i,\sigma}=c^{\dagger}_{i,\sigma}c_{i,\sigma}$, $J_\sigma$ are
the tunnelling amplitudes, and $U_{\sigma}$ are the onsite
interaction energy shifts. In the following, we will consider
balanced densities $\bar{m}_{\sigma}=\bar{n}/3$ for total mean
number density $\bar n$, and in typical realisations we will have
equal tunnelling amplitudes $J_{\sigma}=J$. This model is valid in
the limit $J_\sigma,U_\sigma {\bar n} \ll \omega$, with $\omega$ the
energy separation between the lowest two Bloch bands.

Three-body recombination will result in decay into the continuum of
unbound states, i.e.,  loss from the optical lattice. The decay
dynamics can be described by a master equation in which loss occurs
from a singe site occupied by three atoms at a rate $\gamma_3$
\cite{daleyloss},
\begin{equation}
\dot \rho^{(n)} = - {\rm i}\left(H_{\rm eff} \rho ^{(n)}- \rho ^{(n)}H_{\rm eff}^\dag \right) +\gamma_3 \sum_i  t_i\rho^{(n+3)}t_i^\dag , \nonumber
\end{equation}
where $\rho^{(n)}$ is the system density operator with $n$ atoms,
$t_i = c_{i,1}c_{i,2}c_{i,3}$, and the effective Hamiltonian is
$H_{\rm eff}=H_U - i \gamma_3 t^\dag_{i} t_i/2$. If we begin with an
initial state not involving three body occupation, then via a
mechanism analogous to the quantum Zeno effect, a large loss rate
$\gamma_3 \gg J$ will \textit{suppress} coherent tunnelling that
would produce triply occupied sites. For large $\gamma_3/J$, loss
occurs at an effective rate that decreases as $J^2/\gamma_3$ in
second-order perturbation theory. Then, on a timescale where loss
can be neglected, the system dynamics is described by the
constrained Hamiltonian
\begin{equation}
H_C=\mathcal{P}H_U \mathcal{P},\quad
\mathcal{P}=\prod_{j}\mathcal{P}_j=\prod_{j}(1-m_{j,1}m_{j,2}m_{j,3}),
\nonumber
\end{equation}
where $\mathcal P$ is a projector onto the subspace of states with
at most two atoms per site. We will initially consider the physics
of this model, then return to the master equation in order to test
the assumption that loss is small, and to investigate time-dependent
preparation of states.

Below we focus on the case of a 1D gas, which allows us to make
quantitative predictions for realistic experimental parameters using
time-dependent density matrix renormalisation group (DMRG) methods
\cite{tdmrg} to compute both ground states and near-equilibrium
time-evolution. The 1D case differs from higher dimensions due to
the absence of spontaneous symmetry breaking, but provides a
striking example of the effect of the hard-core constraint in that
an ACS phase appears that is not present in the absence of the
constraint. We first determine a qualitative phase diagram for the
constrained model in the weakly interacting regime using Tomonaga
Luttinger Liquid (TLL) bosonisation techniques, and then go beyond
this limit by using time-dependent DMRG methods to compute the
ground state within the constrained Hamiltonian. We treat both the
SU(3) symmetric case where all interaction constants are equal (as
could be realised, e.g., with alkaline earth atoms
\cite{alkalineearth}), and the case where interactions are unequal
(as are typical in Lithium experiments at low magnetic fields
\cite{lithium}). At the end we combine time-dependent DMRG
techniques with quantum trajectory methods to investigate the
production of an ACS state by computing time evolution for typical
experimental parameters under the full master equation.

\textit{Bosonisation formalism for the constrained Hamiltonian:-} In
order to produce a qualitative phase diagram, we would like to apply
the bosonisation formalism to our system. It is therefore necessary
to implement the constraint in a more explicit way. We have
identified an exact mapping of the constrained fermionic Hamiltonian
$H_C$ to an unconstrained fermionic Hamiltonian which automatically
respects the constraint, at the expense of including higher order
interactions. We introduce projected operators
$d^{\dagger}_{i\sigma}=(\prod_{j\neq i}\mathcal{P}_j)
c^{\dagger}_{i\sigma}$, $d_{i\sigma}=(\prod_{j\neq i}\mathcal{P}_j)
c_{i\sigma}$, entirely in terms of which we express the Hamiltonian.
We verify (i) that the operators $d_{i\sigma}$ obey fermionic
commutations on the subspace where at most two atoms occupy any
site, and that (ii) the Hamiltonian has vanishing matrix elements in
the space with occupations greater than two and (iii) acts as zero
on any state in this latter space. Thus, we arrive at a fermionic
Hamiltonian with built-in constraint, which we analyse with standard
bosonisation techniques. Here we summarise the results, with the
calculations presented in more detail in a forthcoming work
\cite{marcello}. We introduce three bosonic fields
$\phi_{\sigma}(x)$ related to the continuum version of
$(d^{\dagger}_{i\sigma}, d_{i\sigma})$, from which we can construct
a Hamiltonian by taking the linear combinations
$\phi_{c}=(\phi_{1}+\phi_{2}+\phi_{3})/{\sqrt{3}}$, which represents
collective fluctuations of the total density, and
$\phi_{s1}=(\phi_{1}-\phi_{2})/{\sqrt{2}}$ and
$\phi_{s2}=(\phi_{1}+\phi_{2}-2\phi_{3})/{\sqrt{6}}$, which
represent the spin sectors. If we define TLL parameters $K_\alpha$
and conjugate momentum fields $\Pi_\alpha$ corresponding to each
field $\phi_\alpha$ ($\alpha \in \{c,s1,s2 \}$), we then obtain
\begin{eqnarray}
&H&=\sum_{\alpha=c,s1,s2}\left\{\frac{v}{2}[K_{\alpha}\Pi_{\alpha}^{2}+\frac{1}{K_\alpha}(\partial_{x}\phi_{\alpha})^{2}]\right\}+\nonumber\\
&-& \frac{2g_{ss}}{a^2}\cos[\sqrt{2\pi}\phi_{s1}]\cos[\sqrt{6\pi}\phi_{s2}]-\frac{g_{s}}{a^2} \cos[\sqrt{8\pi}\phi_{s1}], \nonumber
\end{eqnarray}
where $v=2aJ \sin[\pi  \bar{n} / 3]$ is the Fermi velocity, $a$
is the lattice spacing. The coefficients $g_{ss}$ and $g_{s}$ exhibit
non-trivial dependence on $U/J$ and ${\overline m}_\sigma$, as do $K_\alpha$.

\textit{Correlation functions:-}  To identify the dominant
correlations and determine a phase diagram, we consider the
behaviour of correlation functions related to different ordering:
CDW (for which we compute density-density correlations
$C(x)\propto\langle n_i n_{i+x}\rangle$), ACS (with BCS correlations
$P_{\sigma}(x)\propto\langle
d^{\dagger}_{i,\sigma}d^{\dagger}_{i,\sigma+1}d_{i+x,\sigma}d_{i+x,\sigma+1}
\rangle $ ) and OT ($OT(x)\propto \langle \tilde
t^\dagger_{i,\sigma}  \tilde t_{i+x,\sigma}\rangle$ with $\tilde
t_{i,\sigma}= d_{i,\sigma}d_{i,\sigma+1}d_{i+1, \sigma+2}$).  Each
correlation is characterized by a scaling dimension that determines
its algebraic decay. The decay exponents for CDW and ACS order are
independent of $\sigma$ in the SU(3) symmetric case and can be
related to TLL parameters as: $\mathcal{D}_{CDW} =
{2K_c}/{3}+{K_{s1}}/{3}+K_{s2}$, $\mathcal{D}_{ACS}=K_{s1}+
{2}/{(3K_c)}+{1}/({3K_{s2})}$. Correlations for OT order have
multiple contributions in terms of bosonic operators, with two
distinct decay exponents in the SU(3) case, $\mathcal{D}^I_{OT}=
{3}/{(2K_c)}+3K_c/2$ and
$\mathcal{D}^{II}_{OT}=3/(2K_c)+K_c/6+4K_{s2}/{3}$. We can then
extract values of the decay exponents as a function of $U/J$ and
$\bar n$ by expanding $K_\alpha$ in the weak interaction limit.
These can be compared to determine a phase diagram, and also
benchmarked against values extracted from numerical simulations (see
Table I).

\textit{Phase diagram from bosonisation (equal interactions):-} The
1D model $H_U$ without the hard-core constraint has been previously
studied for attractive interactions using a combination of the TLL
formalism and DMRG methods \cite{dmrg3}. The qualitative phase
diagram is depicted in Fig.~1a, and involves competition between
symmetric (on-site) trion (ST) order and charge-density-wave (CDW)
order. In a wide region near the SU(3) symmetric line
${U}_{\sigma}=U<0$, CDW is dominant for higher densities and
intermediate interactions. A particular feature of this system with
$U<0$ is that a gap appears in the entire spin sector, so that in
contrast to a two-species Fermi gas, ACS correlations decay
exponentially (as do spin-density wave correlations).

In the presence of the constraint this picture changes substantially
(see Fig.~1b with small $|U/J|$). Outside a small region at low densities $\bar{n}<0.2$
and intermediate interactions, where off-site trionic correlations
compete with CDW, the spin sector is gapless and all correlation
functions decay algebraically. The ST phase is suppressed by the
constraint, and is replaced by a competition between off-site trions
(OT), and an ACS phase with simultaneous pairing of all three pairs
of components.

\textit{Phase diagram from bosonisation (unequal interactions):-} In
the case of unequal interactions, the spin and charge sectors are
coupled by additional terms
$H_{cross}=\lambda_{cs2}\partial_x\phi_{c}\partial_x
\phi_{s}^2+\lambda_{cs1}\partial_x\phi_{c}\partial_x
\phi_{s1}+\lambda_{ss}\partial_x\phi_{s1}\partial_x \phi_{s2}$, with
$\lambda_{cs2}=\sqrt{2}(2U_1-U_2-U_3)/(6\pi v)$, and
$\lambda_{cs1}/\sqrt{6}=(U_3-U_2)/(6\pi v)=-\lambda_{ss}/\sqrt{3}$.
When the interaction imbalance is strong enough to couple different
sectors, opening of a gap in the spin sector can enhance one-channel
pairing with respect to the other, and we obtain pairing between the
species with the strongest attractive interaction.

\begin{figure}[tb]
{\includegraphics[width=8.5cm]{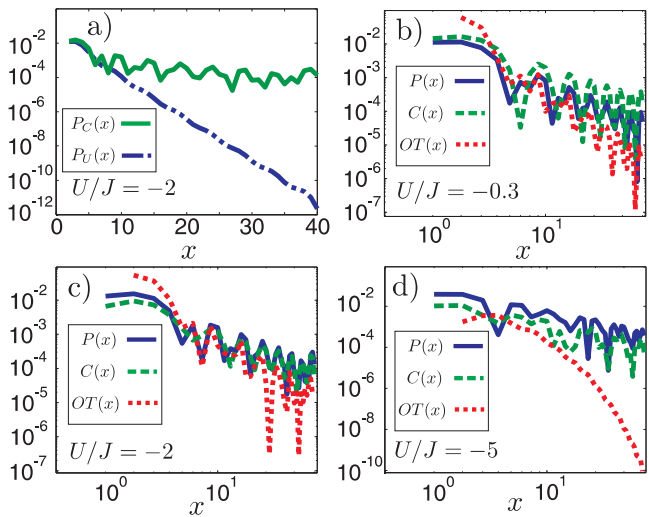}}

\caption{a) ACS-correlations $P_\sigma(x)$ with (solid line) and
without (dash-dotted line) the three-body constraint as a function
of distance $x$ on a 40 site lattice. b) - d) $P_\sigma (x)$ with
CDW-correlations $C(x)$ and off-site trion correlations $OT(x)$ for
different values of $U$, at a density ${\bar m}_1={\bar m}_2={\bar
m}_3=0.2$, with the constraint. In qualitative agreement with
bosonisation, we observe that (b) for weak coupling CDW clearly
dominates. As the coupling increases (c, d), ACS becomes stronger
relative to CDW. For the values presented here, ACS dominates
off-site trions, with $OT(x)$ decaying exponentially for strong
coupling.}

\label{fig:compare}

\end{figure}

\textit{Ground state computations with time-dependent DMRG (equal
interactions) :-} In order to quantitatively underpin these results
and to go beyond the weak-coupling regime, we present calculations
based on time-dependent DMRG methods \cite{tdmrg}. In Fig.~2a, we
see again the striking comparison between the ACS correlations in
the ground states for $H_U$ and $H_C$, which exhibit exponential
decay without the hard-core constraint, and algebraic decay in the
presence of the constraint. In Figs.~2b-c we show a comparison of
the correlations corresponding to the ACS ($P_{\sigma}(x)$), CDW
($C_{\sigma}(x))$ and OT ($OT(x)$), in the ground state of the model
with a three-body hard-core constraint ($H_C$). These are presented
for symmetric but varying interactions from weak to strong coupling,
$U/J\in[-10,-0.3]$. For increasing interactions, the values of
$P_{\sigma}(x)$ become stronger compared with $C_{\sigma}(x)$, so
that the ACS appears to dominate for stronger interactions. While
off-site trions still show algebraic decay for $U/J\geq -2$, they
are subdominant to the ACS and CDW, and they decay exponentially for
$U/J \leq-5$ (Figs. \ref{fig:compare} b-d). This is a strong
deviation from the weak-coupling bosonisation results, and could
indicate an instability (e.g., towards phase separation), or the
appearance of a gap in the dual field of the charge sector. However,
with system sizes of 40 lattice sites, we have not observed any
evidence of phase separation.

Extracting the exponents of the algebraic decay of the correlation
functions further confirms this picture, as shown in Table 1. The constrained model sees an enhancement of
ACS correlations with decreasing $U$, while CDW correlations
generally decay faster as $U$ is lowered. In the weak coupling
regime we also generally observe good agreement with the
perturbative values of the exponents from TLL-theory. In the
strong coupling regime ($U/J \leq -5$) for the constrained case, the
exponents $\mathcal{D}_{ACS}$ and $\mathcal{D}_{CDW}$ saturate, with
$\mathcal{D}_{ACS}$ taking a value compatible with the TLL
prediction $\mathcal{D}_{ACS}^{TLL}=2/(3K_c)+4/3$ in the case where
$2/(3K_c)\rightarrow 0$. This is consistent with either an
instability ($K_c\rightarrow \infty$), or $\Pi_c$ becoming
massive.

\begin{table}
 \centering
 \begin{tabular}{c | c c c }
 \hline
 $-U/J$ & $\mathcal{D}_{ACS}$ & $\mathcal{D}_{CDW}$ & $\mathcal{D}_{OT}$  \\
 \hline
 0.3 & 1.71(0.02)/2.04 & 1.39(0.02)/1.95 & 3.3(0.3)/3.00 \\
 0.6 & 1.66(0.02)/2.03 & 1.40(0.03)/1.96 & 3.3(0.3)/3.00 \\
 1   & 1.60(0.02)     & 1.40(0.03)     & 3.1(0.4)        \\
 2   & 1.4(0.2)     & 1.4(0.1)     & 3.0(0.4)        \\
 5,8,10  & 1.4(0.4)     & 1.4(0.3)     &  exp              \\
 \hline
 \end{tabular}
 \caption{Exponents for algebraic decay of correlations,
computed for ground states in a system of 40 lattice sites with
${\bar m}_\sigma=0.2$ by fitting a power law to the periodic peaks
of the correlations (see Fig.~2). Errors are given in parentheses,
analytic values in the weak coupling limit are given to the right of
slashes. We find agreement with the qualitative predictions of
analytic theory, i.e. for weak coupling we start out with the
CDW-phase dominant, from where $\mathcal{D}_{CDW}$ remains constant
whilst $\mathcal{D}_{ACS}$ and $\mathcal{D}_{OT}$ decrease with
increasing $|U/J|$. We further observe a transition to exponential
decay for the off-site trions in the strong coupling limit.}
 \label{tab:exponents}
\end{table}

\textit{Ground state computations with time-dependent DMRG (unequal interactions):-} In the case of asymmetric interactions we also observe an ACS pairing, but with only the two components paired that exhibit the strongest interactions. As an example, we consider the case of $^6$Li, where in Fig.~3a we plot the Hamiltonian parameters as a function of magnetic field strength near 500-700 G for a fixed lattice depth. From the pairing correlations shown in Fig.~3b, we see that at 520 G, the only algebraically decaying correlations are those corresponding to components 1 and 3, which have the strongest interparticle interaction.

\textit{Time-dependent preparation of states:-} Considering this
example, we now return to the full time-dependent dynamics including
three-body loss, in order to demonstrate a method to produce these
ACS states in $^6$Li. We simulate the many-body master equation on
12-24 lattice sites by combining time-dependent DMRG methods with
quantum trajectories techniques, as described in \cite{daleyloss}.
We assume that the lattice is initially loaded at a magnetic field
of $615$ G, where the repulsive interactions (see Fig.~3a) will
stabilise the system in the presence of loss. We then consider  a
time-dependent ramp of the magnetic field to $500$ G. The
characteristics of the ramp we choose (shown in Fig.~3c) are: (i) it
is adiabatic until $565$ G, where the components 2 and 3 become
paired, (ii) it is fast from $565-500$ G, where onsite trions become
energetically favoured ($\sum_{\alpha}U_{\alpha}<U_{2}$) and where
for fields larger than $520$ G, $\gamma_3$ is too small to prevent
triple occupation (see Fig.~3a), and (iii) after a hold time
$T_{h,1}=16J^{-1}$ we add a swap between species 1 and 2 via a fast
laser pulse at the end of the ramp, after which there is a second
hold time $T_{h,2}=20J^{-1}$. The ACS correlations in the final
state after the swap (Fig.~3d) then exhibit dominant pairing between
species 1 and 3, as would be expected in the ground state for the
parameters of $^6$Li at $B=500$ G with the constraint. At the same
time, the other two pairing channels are clearly subdominant as
evidenced by their much weaker correlators, while trion formation is
also strongly suppressed during the ramp. Further, the probability
that not even a single decay event occurs during the ramp shown here
is $\sim 80$\% for 12 lattice sites.

These ideas can be extended to higher dimensions, where the ACS should also
be stabilised due to suppressed trion formation.

\begin{figure}[tb]

\begin{center}
\includegraphics[width=8.5cm]{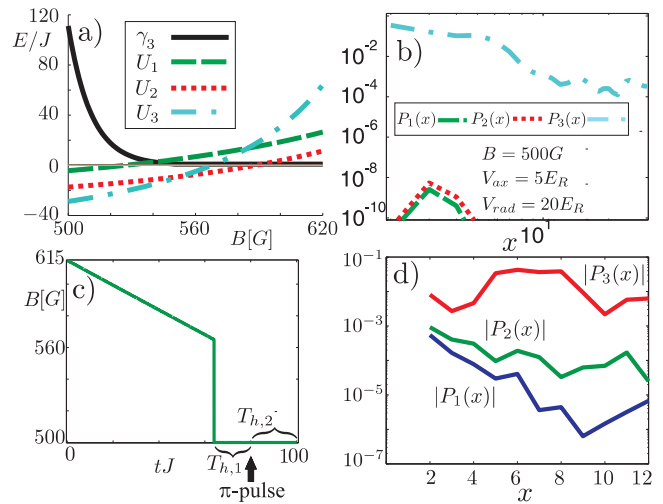}
\end{center}

\caption{a) Hubbard parameters $U_1$, $U_2$, $U_3$ and three-body
loss rate $\gamma_3$ for $^6$Li as function of external magnetic
field $B$, for a lattice depth of $V_{ax}=5$ $E_R$ in axial
direction, and $V_{rad}=20$ $E_R$ in radial direction. b) ACS
correlations for the ground state of $H_C$ at $B=500$ $G$, computed
for a system of $30$ lattice sites, ${\bar m}_\sigma=2/15$,
$\sigma=1,2,3$. The solid line shows $P_3(x)$, dashed line $P_1(x)$,
dotted line denotes $P_2(x)$. As $U_3<U_2\ll U_1$, we see $1-3$
pairing dominate, with all other pairing exponentially suppressed.
c) Magnetic field $B(t)$ for the adiabatic ramp. d) ACS correlations
after the time-dependent ramp of the magnetic field shown in c),
beginning from the ground state at $615$ G, with $\bar m_\sigma =
0.16$, on $12$ sites.} \label{fig:Li}
\end{figure}

\begin{acknowledgments}
We thank S. Jochim, M. Fleischhauer, P. Julienne, M.
Baranov, and E. Ercolessi for discussions. This work was
supported by the Austrian Science Foundation (FWF) through SFB F40
FOQUS and project I118\_N16 (EuroQUAM\_DQS), the DARPA OLE program,
and by the Austrian Ministry of Science BMWF via the
UniInfrastrukturprogramm of the Forschungsplattform Scientific
Computing and Centre for Quantum Physics Innsbruck.
WH acknowledges support by the DFG through SFB/TR 49.
\end{acknowledgments}

\end{document}